\documentclass[a4paper,11pt]{article}
\usepackage{pos}
\usepackage{float} 
\usepackage{multirow} 
\usepackage{subfig} 

\title{Study of acoustic neutrino detection in O$\nu$DE-2 raw acoustic data}
\author[a]{D. Bonanno}
\author[a]{L. S. Di Mauro}
\author*[a]{D. Diego-Tortosa}
\author[a]{A. Idrissi}
\author[a]{G. Riccobene}
\author[a]{S. Sanfilippo}
\author[a]{S. Viola}

\affiliation[a]{Istituto Nazionale di Fisica Nucleare (INFN) - Laboratori Nazionali del Sud (LNS),\\
  Via S. Sofia 62, Catania, 95123, Italy}

\emailAdd{didac.diegotortosa@lns.infn.it}

\abstract{Research suggests that acoustic technology may be able to detect ultra-high-energy neutrinos if a large amount of non-linear fluid is analyzed. When a neutrino interacts in water, it creates a quasi-instantaneous cascade of particles, heating that region of the fluid and emitting a tiny acoustic signal. This rapid heating produces a thermoacoustic Bipolar Pulse (BP) with unique characteristics such as a wide bandwidth and a narrow directivity for these frequencies. While dedicated devices for acoustic neutrino detection are currently non-existent, there are a few underwater neutrino telescopes that utilize optical technology, but often with an acoustic positioning system that deploys hydrophones in the infrastructure. The possibility of using them to study a BP caused by a neutrino interaction is currently being discussed. This study aims to evaluate the implementation of a trigger system to detect a possible BP in deep-sea hydrophones. For this, up to 24 hours of the raw acoustic signal recorded by the O$\nu$DE-2 station, which was located 25 km off-shore from Catania in the Western Ionian Sea, at 2100 m depth, is analyzed. The station used calibrated hydrophones from a few Hz to 70 kHz. In this work, a synthetic BP is created and added to the experimental data, allowing the study of its detection and the calculation of precision and recall.}

\FullConference{10th International Workshop on Acoustic and Radio EeV Neutrino Detection Activities (ARENA2024)\\
11-14 June 2024\\
The Kavli Institute for Cosmological Physics, Chicago, IL, USA\\}

\renewcommand{\logo}{\relax}

\usepackage[style=numeric-comp, sortcites, maxcitenames=3, maxbibnames=3, doi=true, url=false, sorting=none]{biblatex} 
\bibliography{ARENA24bib}

\begin{document}
\maketitle

\section{Introduction}
    The main objective of this work is to test the acoustic trigger alert for neutrino detection, which was presented during the last ARENA conference, discussed in \cite{Diego2022ARENA}. Given that this software is sensitive to marine mammals' echolocation clicks, this study aims to evaluate its performance in a more challenging environment to assess its limitations. The data analyzed correspond to 24H continuous recordings made on 26 February 2017 by one of the hydrophones on the Ocean Noise Detection Experiment (O$\nu$DE). This paper describes the concept of evaluating the trigger by directly inserting three different types of bipolar pulses that mimic the acoustic signature of a neutrino interaction in water, details its configuration and implementation, and presents the results.  
    
    \subsection{The SMO-O$\nu$DE-2 observatory}
        The Submarine Multidisciplinary Observatory (SMO) project funded the construction of a marine station designed for long-term monitoring of the deep-sea environment in O$\nu$DE. This station, called SMO-O$\nu$DE2 (O$\nu$DE-2), was used in a test site before the construction of the NEutrino Mediterranean Observatory (NEMO) Phase-II. 

\begin{figure}[h]
        	\centering
        	\subfloat[]{\includegraphics[height=4.5cm]{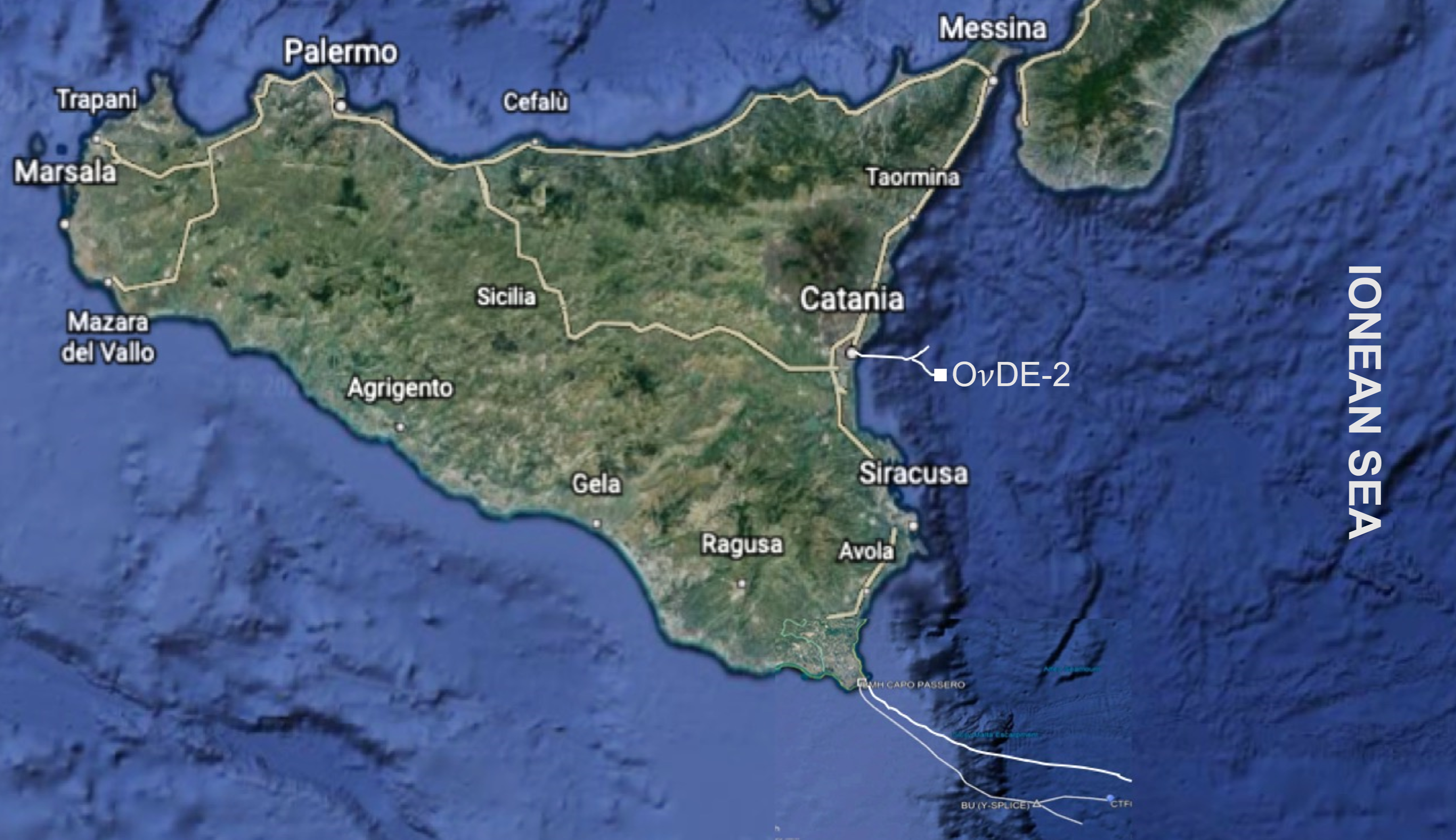}}
        	\hspace{0.7cm}
        	\subfloat[]{\includegraphics[height=4.5cm]{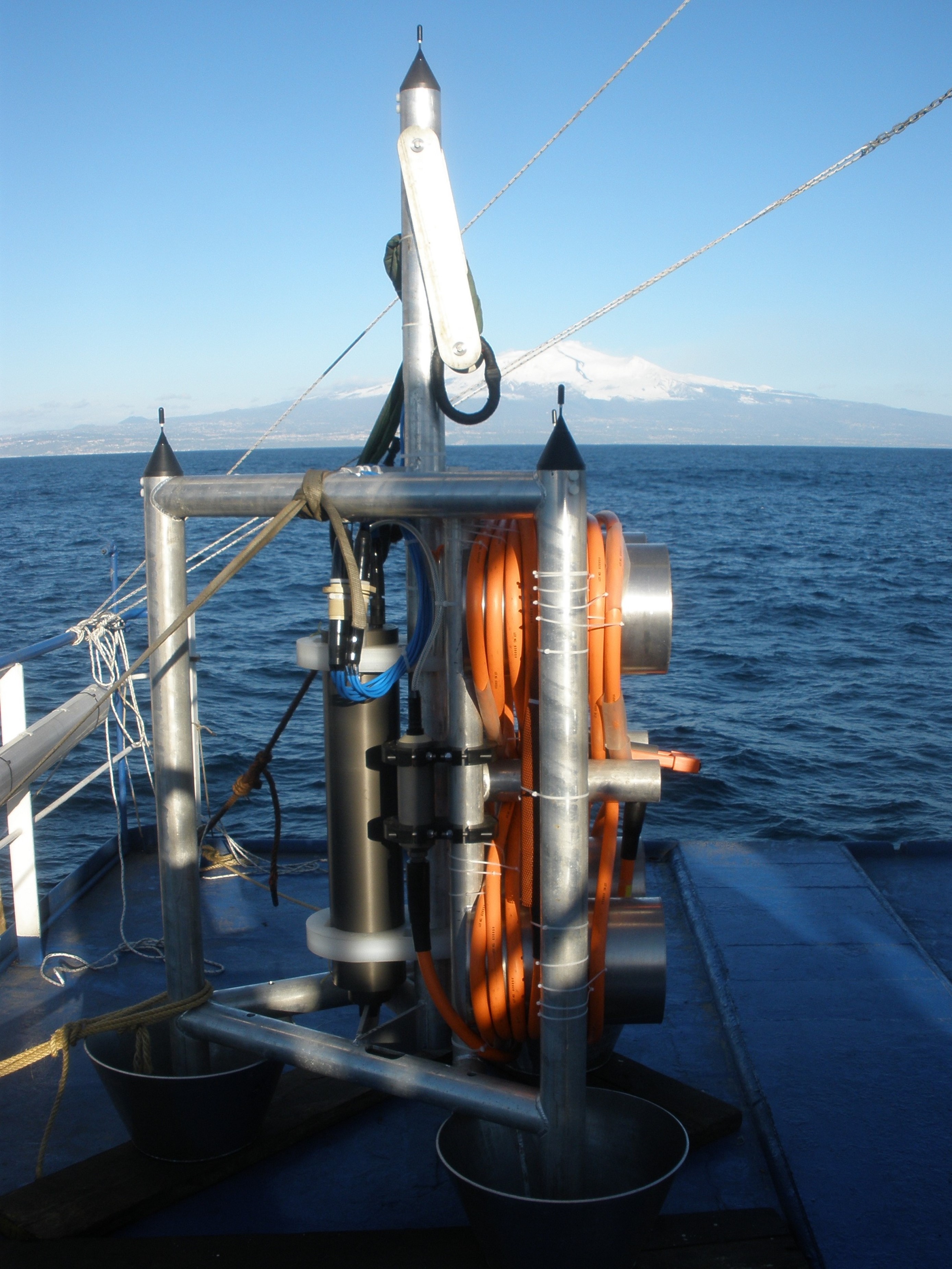}}
        	\caption{\textbf{(a)} Map of Sicily showing the submarine cables and the location of O$\nu$DE-2. \textbf{(b)} The mechanical frame hosting O$\nu$DE-2 acoustic array.}
        	\label{fig:OnDE2}
        \end{figure}

        It has been fully operational for over four years (from February 2017 to May 2021), connected to the Test Site South (TSS) termination of the Main Electro Optical Cable (MEOC) operated by INFN-LNS. This site is located 25 km offshore from Catania at a depth of 2100 meters and is part of the Western Ionian Sea within the European Multidisciplinary Seafloor and water-column Observatory (EMSO) project (see \textit{Fig.\ref{fig:OnDE2}}). Among other sensors, it features four SMID TR-401 hydrophones arranged in a tetrahedral pyramidal shape, with a distance of 1 meter between them (sensitivity of -172$\pm$3 dB re 1 V/$\mu$Pa in a 10 Hz--70 kHz frequency range \cite{Viola2013}). The hydrophones are synchronized by the same GPS signal clock, and their data acquisition system comprises two stereo Analog-to-Digital Converter boards with a sampling frequency of 192 kHz and a resolution of 24 bits \cite{Viola2018ARENA}.
        
    \subsection{The acoustic neutrino Bipolar Pulse}
        When a muon neutrino ($\nu_{\mu}$) interacts with water or ice, it generates a thermo-acoustic Bipolar Pulse (BP) \cite{Diego2022ARENA}. There are several equations to simulate a BP. In this work, the generation of a $BP(t)$ has been simplified by deriving it from a normal distribution $g(t)$, as in \textit{Eq.\ref{eq:GNeq}}. 

        \begin{equation}\label{eq:GNeq}
        	g\left( t \right)=e^{-\frac{1}{2}\left( \frac{t}{\sigma} \right)^2}=e^{\frac{-t^2}{2\sigma ^2}}
        \end{equation}
        where the $\sigma$ represents the standard deviation at a confidence interval of $\sim$68\%. 
        
        To control the width of the BP, the inter-peak value $\Lambda$ should be equivalent to $2\sigma$, as in \textit{Eq.\ref{eq:BPeq}}. 
        
        \begin{equation}\label{eq:BPeq}
        	BP\left(t\right)=\frac{dg}{dt}=-\frac{t}{\sigma^2}\cdot e^{\left[\frac{-t^2}{2\sigma^2}\right]} \quad {}^{\underrightarrow{\ \Lambda=2\sigma \hphantom{\sigma}}} \quad BP\left(t\right)=\frac{-4t}{\Lambda ^2}\cdot e^{-2\ \left(\frac{t}{\Lambda}\right)^2}
        \end{equation}
        
\begin{figure}[h]
        	\centering
        	\subfloat[]{\includegraphics[height=4cm]{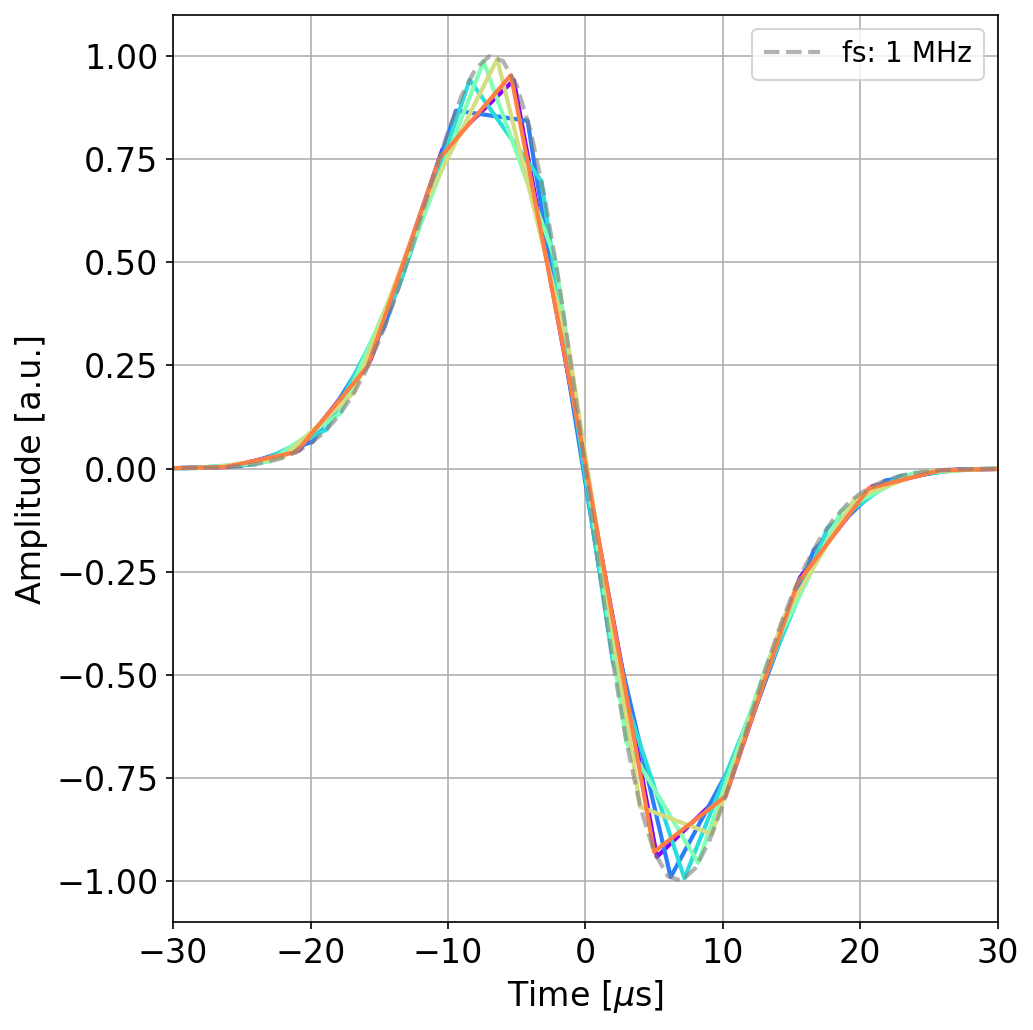}}
        	\hspace{0.7cm}
        	\subfloat[]{\includegraphics[height=4cm]{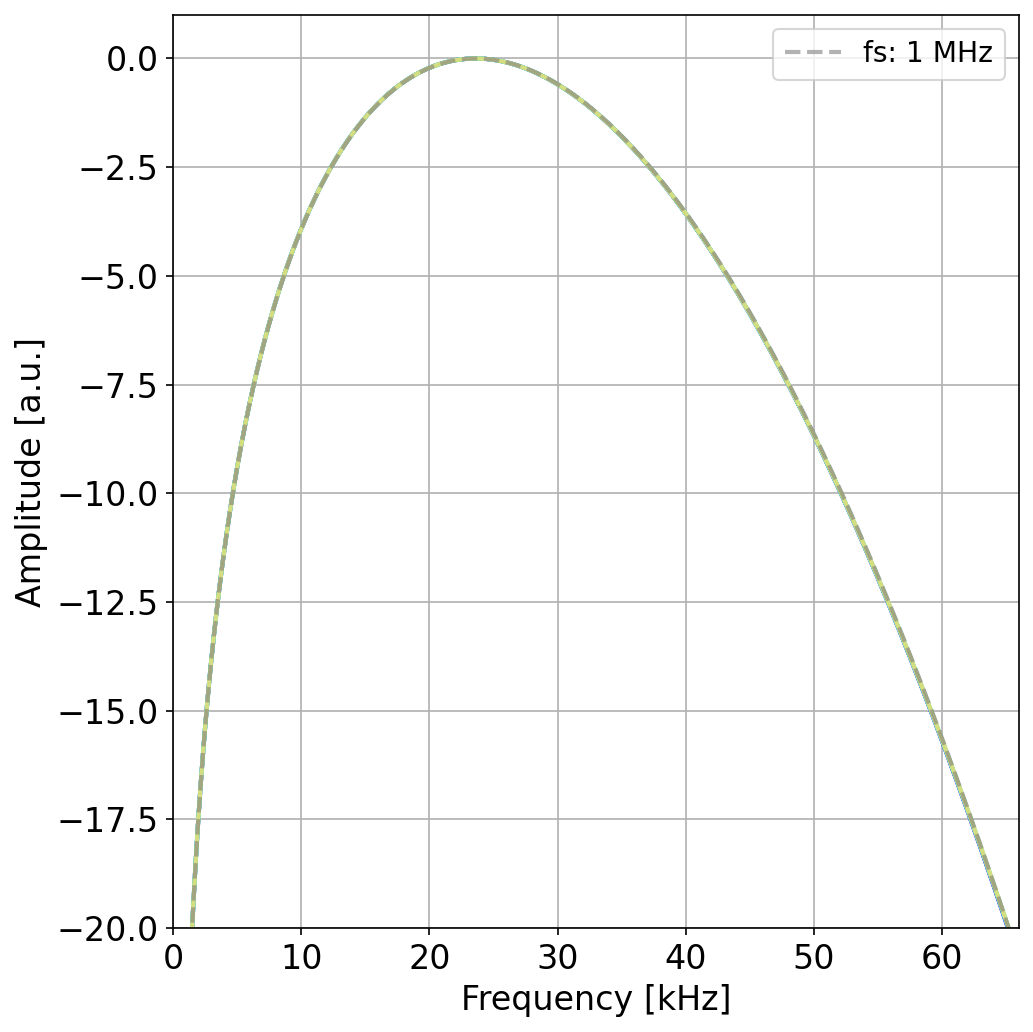}}
        	\caption{\textbf{(a)} Generated BPs (colored) to be inserted into the experimental data, with an inter-peak distance value ($\Lambda$) of 12.5 $\mu$s. \textbf{(b)} Frequency domain of the created BPs.}
        	\label{fig:BP}
        \end{figure}

        \textit{Fig.\ref{fig:BP}} shows the result of the BP that will be inserted into the raw acoustic data representing a neutrino interaction observed at 1 km and 0-degree orientation from the source. The corresponding signal was generated with a sampling frequency of 1 MHz, and from this, six different possible BPs have been created using the sampling frequency used by O$\nu$DE-2  (192 kHz). These will be randomly selected for embedding into the experimental signals.

        These simulated BPs, which correspond to the "true" event for the evaluation, will be used to represent a neutrino deposition of 10$^{10}$, 10$^{11}$, and 10$^{12}$ GeV. These energies correspond to maximum peak pressures of 12.4 mPa, 123.9 mPa, and 1.2 Pa, respectively \cite{Bevan2009}.

        The BP used in this study exhibits significant similarities to bioacoustic echolocation pulses (clicks), which can lead to potential confusion in signal analysis. The six signals contain energy starting at 1.48, 3.68, 7.59, and 11.41 kHz with -20, -12, -6, and -3 dB decays, respectively, and extending up to 65.18 $\pm$ 0.0266 kHz and 55.08 $\pm$ 0.0136 kHz at -20 and -12 dB decays. This wide frequency range makes it difficult to identify the exact spectral characteristics. Similarly, echolocation clicks produced by marine mammals, such as sperm whales and dolphins, are characterized by very short pulses, typically around 100 $\mu$s for sperm whales and between 40 and 110 $\mu$s for some dolphin species \cite{Mhl2003,deFreitas2015}. These clicks, being signals similar to BP in time, also exhibit a large bandwidth, commonly within the range discussed for the BPs. Both signals present a wide bandwidth over a brief duration, which complicates the acoustic detection of neutrinos \cite{Diego2022Thesis}.
        
\section{The BP trigger alert}
\begin{figure}[htbp]
		\centering
		\includegraphics[height=4cm]{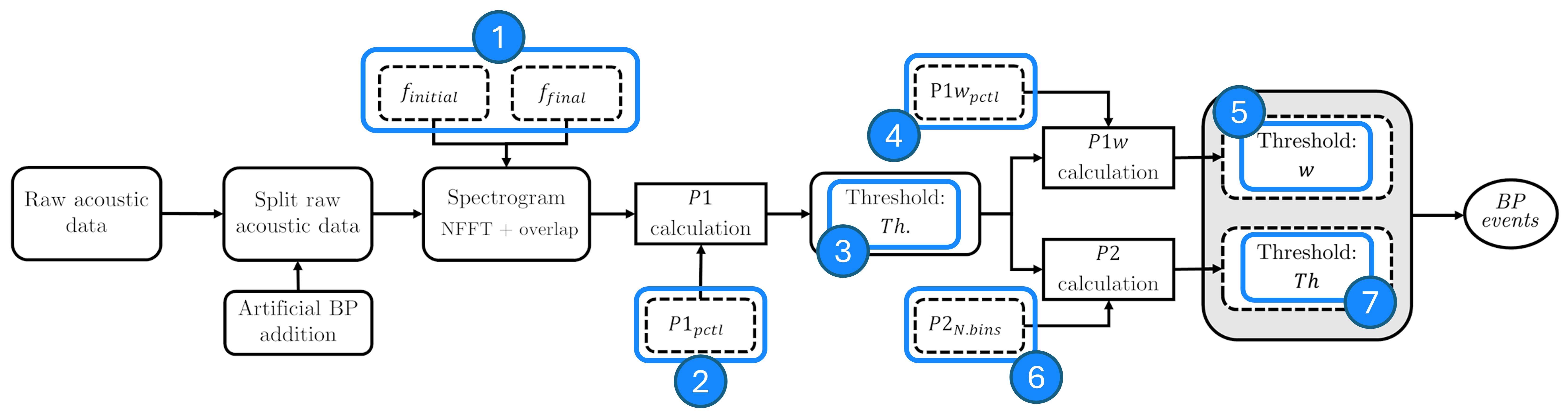}
		\caption{Workflow of the BP trigger (1st level of the trigger: analysis per receiver).}
		\label{fig:Workflow}
	\end{figure}

    This preliminary study is focused on the application and analysis of results only at the first trigger level of the acoustic neutrino detector \cite{Diego2022ARENA}, using data from a single hydrophone (see \textit{Fig.\ref{fig:Workflow}}). If the second-level trigger is applied, which involves selecting coincident events recorded by different hydrophones, no significant changes in the results are expected. Given the one-meter distance between the tetrahedron-shaped O$\nu$DE-2 hydrophones and the synchronization accuracy, the directivity of both BP and bioacoustic clicks may not be differentiated among them. This makes it difficult to account for directivity when selecting BP clicks.
    
    The raw acoustic data are split into one-second splits, and an artificial BP is randomly added in these segments, random in terms of the BP selection and its temporal insertion. A spectrogram is then generated for each segment, different parameters are calculated, and up to three cuts are applied to select the found BP events.
    
    \subsection{The NFFT spectrogram selection}
        The spectrogram was configured with an overlap of 50$\%$ to meet three objectives: the time resolution $T_{\text{res.}}$ should be less than 100 $\mu$s, the frequency resolution $F_{\text{res.}}$ should be less than 5 kHz and the lowest valid frequency $F_{\text{min.}}$ should be less than 10 kHz. The 38-NFFT stands out as the optimal choice to fulfill the three specified objectives. Despite not being a base-two value, it is selected for its superior performance (see \textit{Tab.\ref{tab:Tab1}}).
        
        \begin{table}[H]
            \centering
            \begin{tabular}{cccc}
            \hline
            \multirow{2}{*}{NFFT} & $T_{\text{res.}}$ [$\mu$s]                    & $F_{\text{res.}}$ [kHz]                    & $F_{\text{min.}}$ [kHz]                  \\
                                  & Objective $\lesssim$ 100 & Objective $\lesssim$ 5 & Objective $\lesssim$ 10 \\ \hline
            32                    & 83.3                  & 6.0                     & 12.0                    \\
            \textbf{38}                    & \textbf{99.0}                    & \textbf{5.1}                     & \textbf{10.1}                    \\
            64                    & 166.7                   & 3.0                     & 6.0                    \\ \hline
            \end{tabular}
            \caption{\label{tab:Tab1}Spectrogram parameters depending on the selected NFFT.}
        \end{table} 
    
    \subsection{The tuning cut-offs}\label{subsec:TuningCutOffs}
    Since the environment and sensitivity of the hydrophones influence the characteristics of the recorded signals, the trigger parameters must be adjusted. The parameters to be adjusted are as follows:
    \begin{enumerate}
        \item[(1)] $f_{initial}$ and $f_{final}$: Frequency range where the BP energy is to be searched. 
        \item[(2)(3)] \textit{Cutoff} $P1$: $P1$ is the parameter controlling the event energy is the average Power Spectral Density (PSD) value within the frequency range of interest. This parameter is set dynamically using a chosen percentile value ($P1_{pctl}$) combined with an additional absolute threshold in dB ($P1_{th}$). Together, they determine this first selection cutoff.
        \item[(4)(5)] \textit{Cutoff} $P1w$: The parameter controlling the duration of the event corresponds to the width of an event that passes the cutoff $P1$ and is evaluated by considering the time during which the peak exceeds a percentile $P1w_{pctl}$ of its maximum. Then, $P1_{th}$ is used to limit the selection of possible BP events and discard long signals, establishing the duration cutoff.
        \item[(6)(7)] \textit{Cutoff} $P2$: $P2$ is the difference between the $P1$ value of a the sample candidate (in a single spectrogram bin) and the mean $P1$ value of the surrounding samples. $P2_{N.bins}$ determines the number of surrounding bins to consider in the $P2$ calculation. Using $P2_{th}$, a minimum dB value is selected. Therefore, it is another parameter used to control the duration of the event.
    \end{enumerate} 
    
    To adjust these trigger parameters, it is advisable to perform a preliminary study with samples of the experimental data by testing different values and adjusting them according to the trigger's objectives. In this study, 25$\%$ of the data has been used for this purpose, specifically the last hour of each set of four.
    In this test, the $f_{initial}$ parameter was set at 10 kHz while the $f_{final}$ was tested between 45 and 65 kHz in 5 kHz steps. For $P1_{pctl}$, the 80th, 85th, 90th, 95th, 98th, and 99th percentiles were tested, while the threshold $P1_{th}$ was varied between 3 and 5 dB. For $P1w_{pctl}$, the 75th, 80th, and 90th percentiles were tested, while the threshold $P1w_{th}$ was set with different values between 170 and 300 $\mu$s. Finally, $P2_{N.bins}$ was set to 50, and thresholds $P2_{th}$ of 5 and 7 dB were tested.
    
    \subsection{The trigger evaluation}
    Since a simulated BP event will be added directly to the raw acoustic data, this allows for the study of its detectability. The artificial BP represents a True Positive ($TP$), while all other detections are considered False Positives ($FP$). Additionally, any true BP events that are missed by the algorithm are categorized as False Negatives ($FN$). The performance of the algorithm can be evaluated using common metrics such as precision (accuracy) and recall (the ability of the model to capture all true BP events), as shown in \textit{Ec.\ref{eq:Precision}} and \textit{Ec.\ref{eq:Recall}}.

        \begin{equation}\label{eq:Precision}
			Precision = \frac{TP}{TP+FP}
		\end{equation}
		\begin{equation}\label{eq:Recall}
			Recall = \frac{TP}{TP+FN}
		\end{equation} 

    Additionally, the F1-score, which complements both precision and recall, can be calculated as shown in \textit{Ec.\ref{eq:F1_score}}.
  
        \begin{equation}\label{eq:F1_score}
            F1-score = 2 \cdot \frac{Precision \cdot Recall}{Precision + Recall}
        \end{equation}
  
\section{Results}
\begin{figure}[h]
        	\centering
        	\subfloat[]{\includegraphics[height=4.5cm]{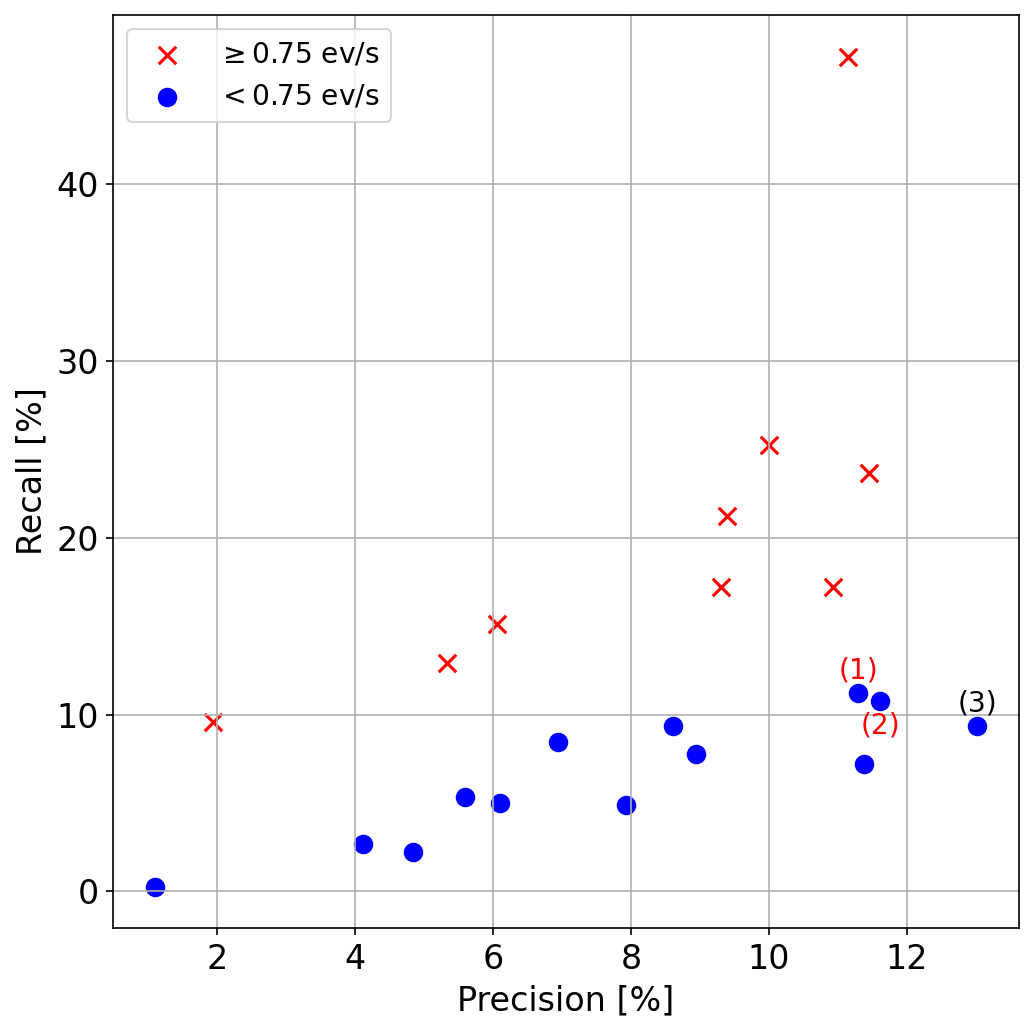}}
        	\hspace{0.7cm}
        	\subfloat[]{\includegraphics[height=4.5cm]{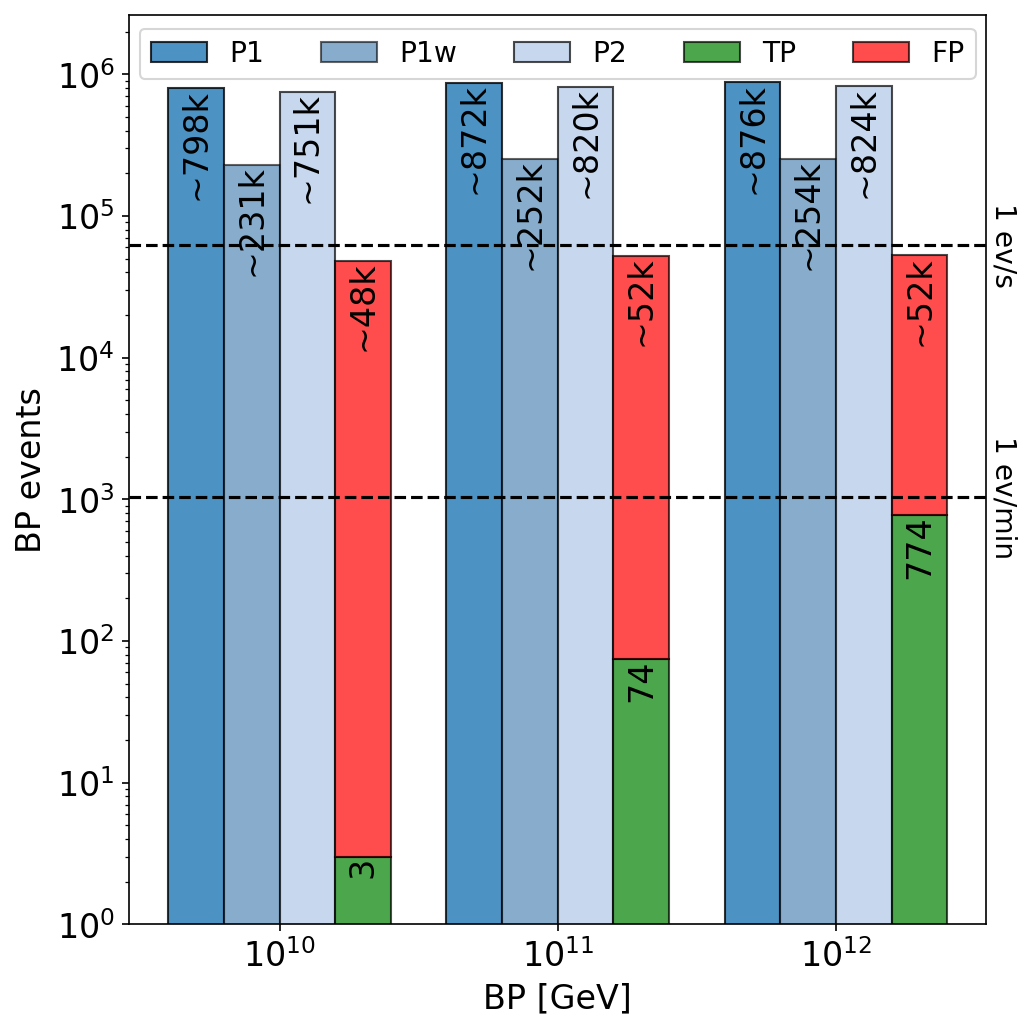}}
        	\caption{\textbf{(a)} Precision and Recall of the different trigger configurations tested with 25$\%$ of the experimental data. The three with the highest F1-score meeting the 0.75 ev/s constraint are listed. \textbf{(b)} Number of events selected in the full data for each cutoff according to the energy of the inserted BP. The last bar shows the number of BP events that passed all three cutoffs.}
        	\label{fig:ResultsPre}
        \end{figure}

As explained in \textit{Section 2.2}, the first step was to find the best trigger parameter values. This test was performed with a BP of 10$^{11}$ GeV each second of analyzed data. To keep the number of detected events below one per second (as more would result in saving excess data), in this test the number of events was limited to 0.75 ev/s (it is hoped that the number of detected events will increase when analyzing the full dataset).
\textit{Fig.\ref{fig:ResultsPre}.a} shows the results of this test with various configurations. The top three F1-score cases that meet the 0.75 ev/s constraint are enumerated.

    The total data analyzed corresponds to almost 18 hours (62397 s), despite some errors in the Data Acquisition (DAQ) system. The BPs were added one per minute.
    Finally, the first two configurations were discarded because, although they stayed below 0.5 ev/s during testing, they exceeded the ev./s rate in the final dataset. Therefore, the third and final configuration presented in this work (\textit{Fig.\ref{fig:ResultsPre}.b}) study the energy in the frequency range of 10 to 60 kHz, using the 98th percentile with a 3 dB threshold for $P1$ calculation, the 80th percentile with a limit of 200 $\mu$s for $P1w$, and considering 50 samples on each side of the event with a 5 dB threshold for $P2$. \textit{Tab.\ref{tab:TabRes}} shows the results with this latter configuration.

\begin{table}[htbp] 
\centering 
\begin{tabular}{ccccccccc}
\hline
\textbf{\begin{tabular}[c]{@{}c@{}}BP\\ {[}GeV{]}\end{tabular}} & \textbf{\begin{tabular}[c]{@{}c@{}}Artificial\\ BPs\end{tabular}} & \textbf{\begin{tabular}[c]{@{}c@{}}Detections\\ {[}ev./s{]}\end{tabular}}& \textbf{\begin{tabular}[c]{@{}c@{}}TP\end{tabular}} & \textbf{\begin{tabular}[c]{@{}c@{}}FP\end{tabular}} & \textbf{\begin{tabular}[c]{@{}c@{}}FN\end{tabular}} & \textbf{\begin{tabular}[c]{@{}c@{}}Precision\\ {[}\%{]}\end{tabular}} & \textbf{\begin{tabular}[c]{@{}c@{}}Recall\\ {[}\%{]}\end{tabular}} & \textbf{\begin{tabular}[c]{@{}c@{}}F1-score\\ {[}\%{]}\end{tabular}} \\ \hline
$10^{10}$                                                              & \begin{tabular}[c]{@{}c@{}}31199\\ {\color[HTML]{FE0000}(30/min)}\end{tabular}            & \begin{tabular}[c]{@{}c@{}}0.7667\end{tabular}         & 3                                                                   & 47839                                                                  & 31196                                                                    & 0.0063                                                                  & 0.0096     
& 0.01\\
$10^{11}$                                                              & \begin{tabular}[c]{@{}c@{}}1023\\ (1/min)\end{tabular}            & \begin{tabular}[c]{@{}c@{}}0.8383\end{tabular}         & 74                                                                    & 52232                                                                  & 949                                                                    & 0.14                                                                 & 7.23       &0.28                                                        \\
$10^{12}$                                                            & \begin{tabular}[c]{@{}c@{}}1023\\ (1/min)\end{tabular}          & \begin{tabular}[c]{@{}c@{}}0.8496\end{tabular}        & 774                                                                     & 52241                                                                  & 249                                                                  & 1.46                                                                & 75.66    &2.86                                                         \\ \hline
\end{tabular}
\caption{Results of performance metrics for the different BP energy}
\label{tab:TabRes}
\end{table}

The 10$^{12}$ GeV BP is 75.66$\%$ detectable of the cases (recall), while the 10$^{11}$ GeV BP is only in the 7.23$\%$. Finally, the 10$^{10}$ GeV BP was not discovered until the appearance rate increased to 0.5/s, and after analyzing the 18 hours of data, only three instances were detected. This suggests that the detections were random.

\begin{figure}[htbp]
		\centering
		\subfloat[]{\includegraphics[width=4.5cm]{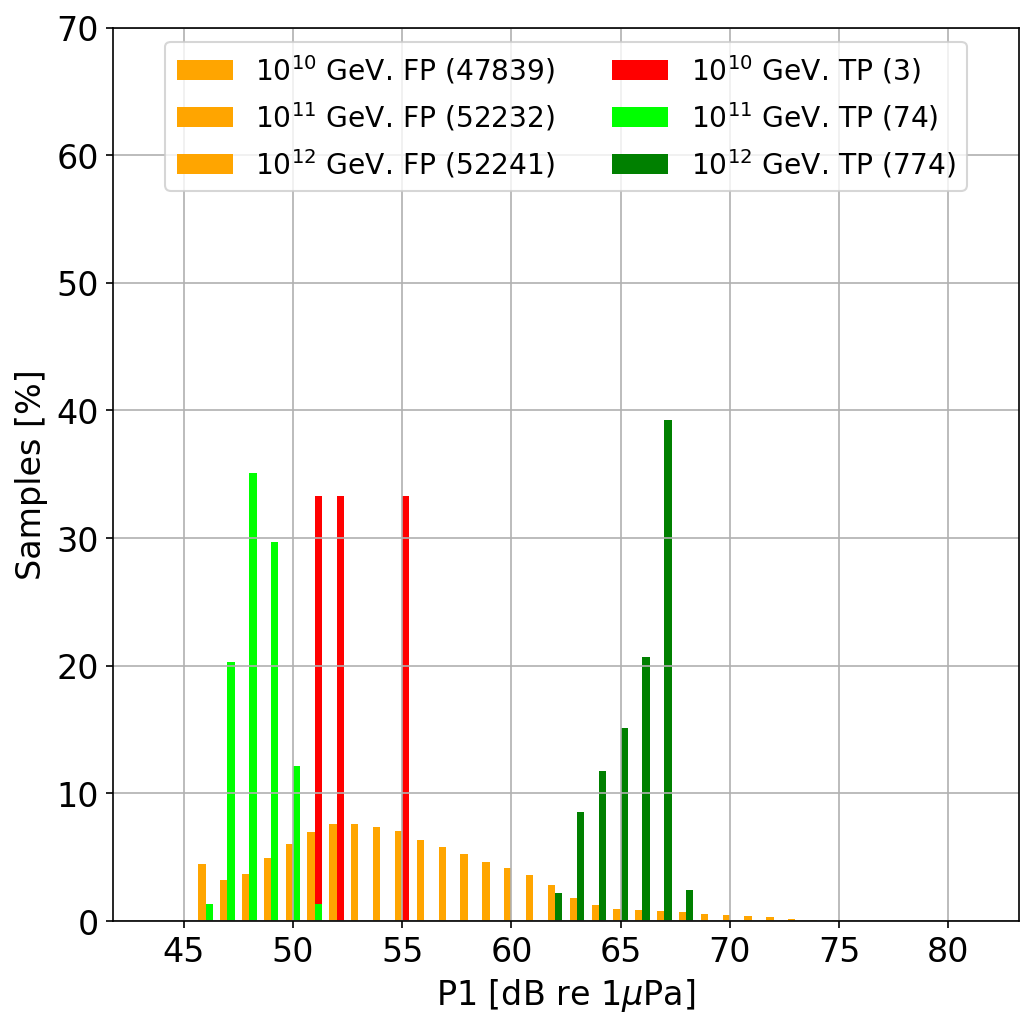}}
		\hspace{.5cm}
		\subfloat[]{\includegraphics[width=4.5cm]{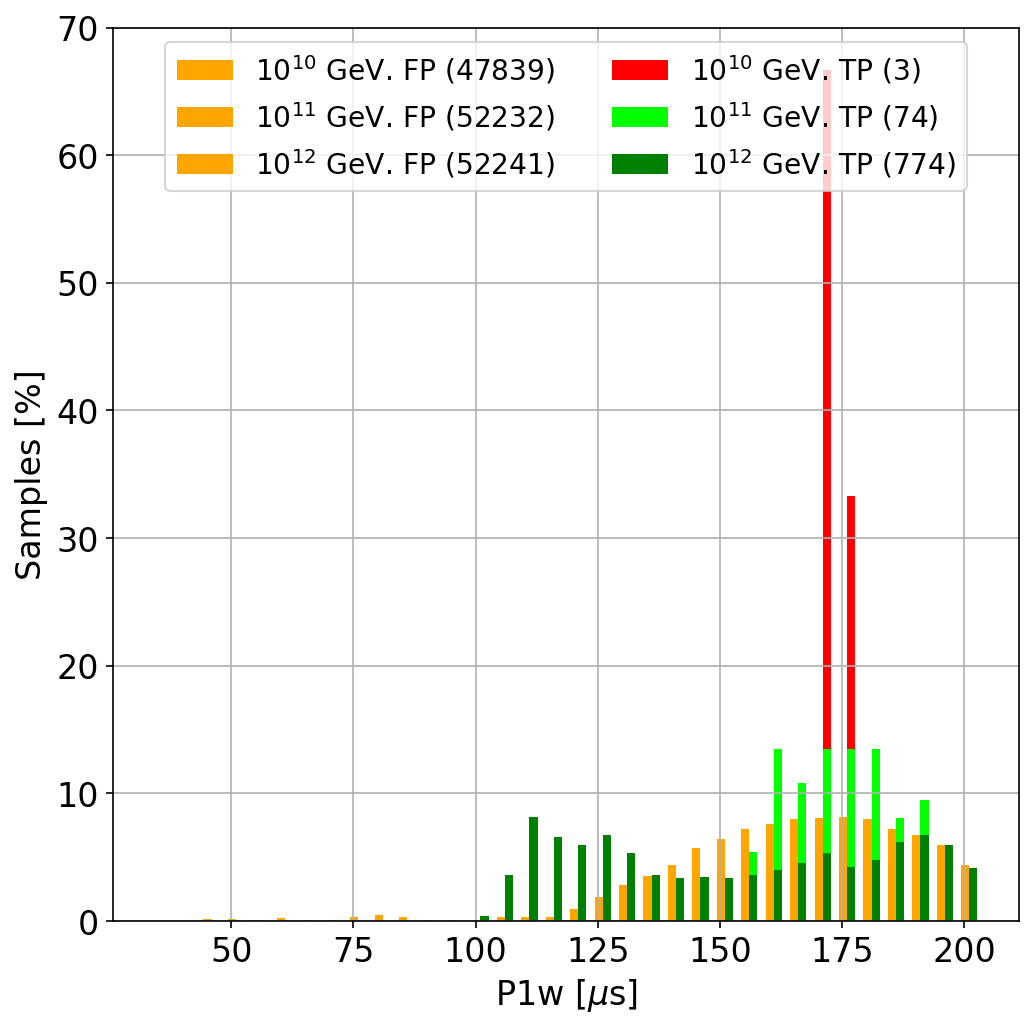}}
		\hspace{.5cm}
		\subfloat[]{\includegraphics[width=4.5cm]{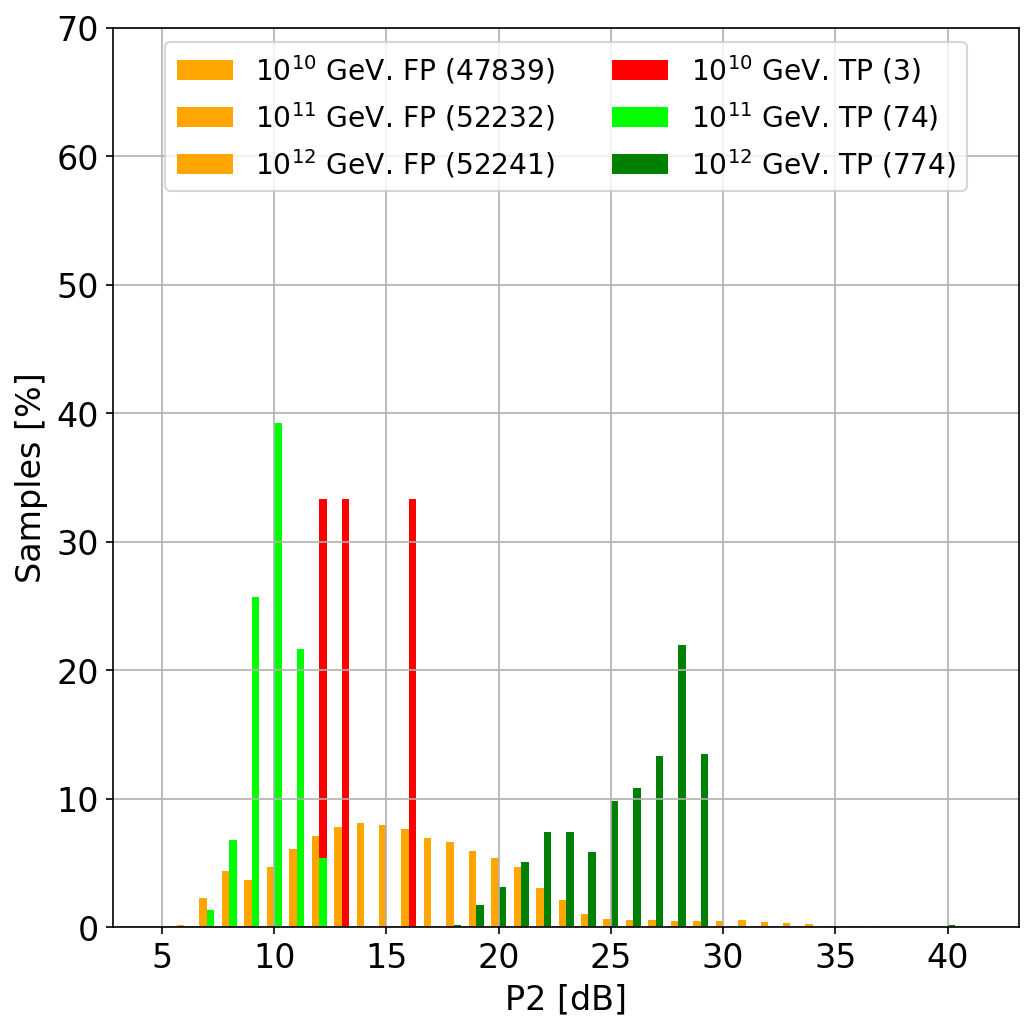}}
		\caption{Percentage of TPs and FPs according to the value of: \textbf{(a)} $P1$, \textbf{(b)} $P1w$, and \textbf{(c)} $P2$, for the three different BP energies.}
		\label{fig:ResGraph1}
	\end{figure}

\begin{figure}[htbp]
		\centering
		\subfloat[]{\includegraphics[width=4.5cm]{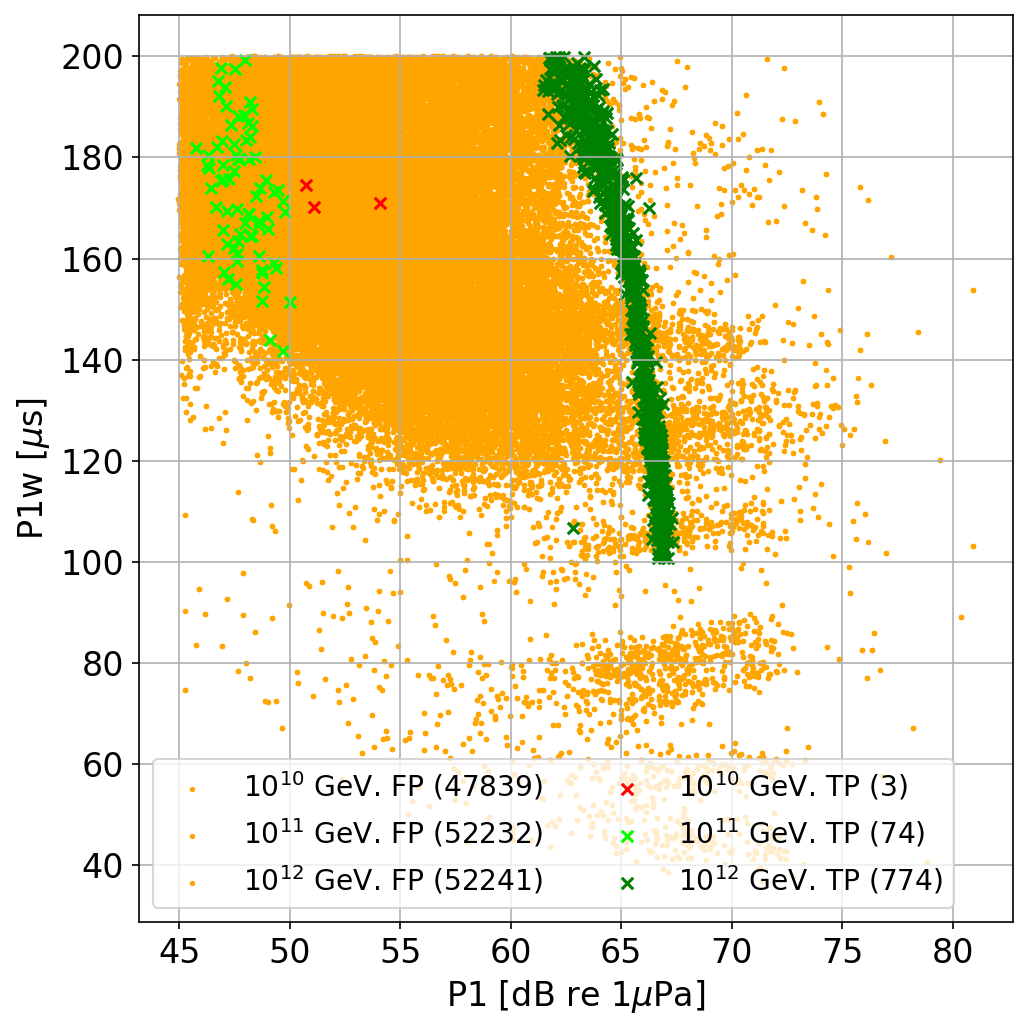}}
		\hspace{.5cm}
		\subfloat[]{\includegraphics[width=4.5cm]{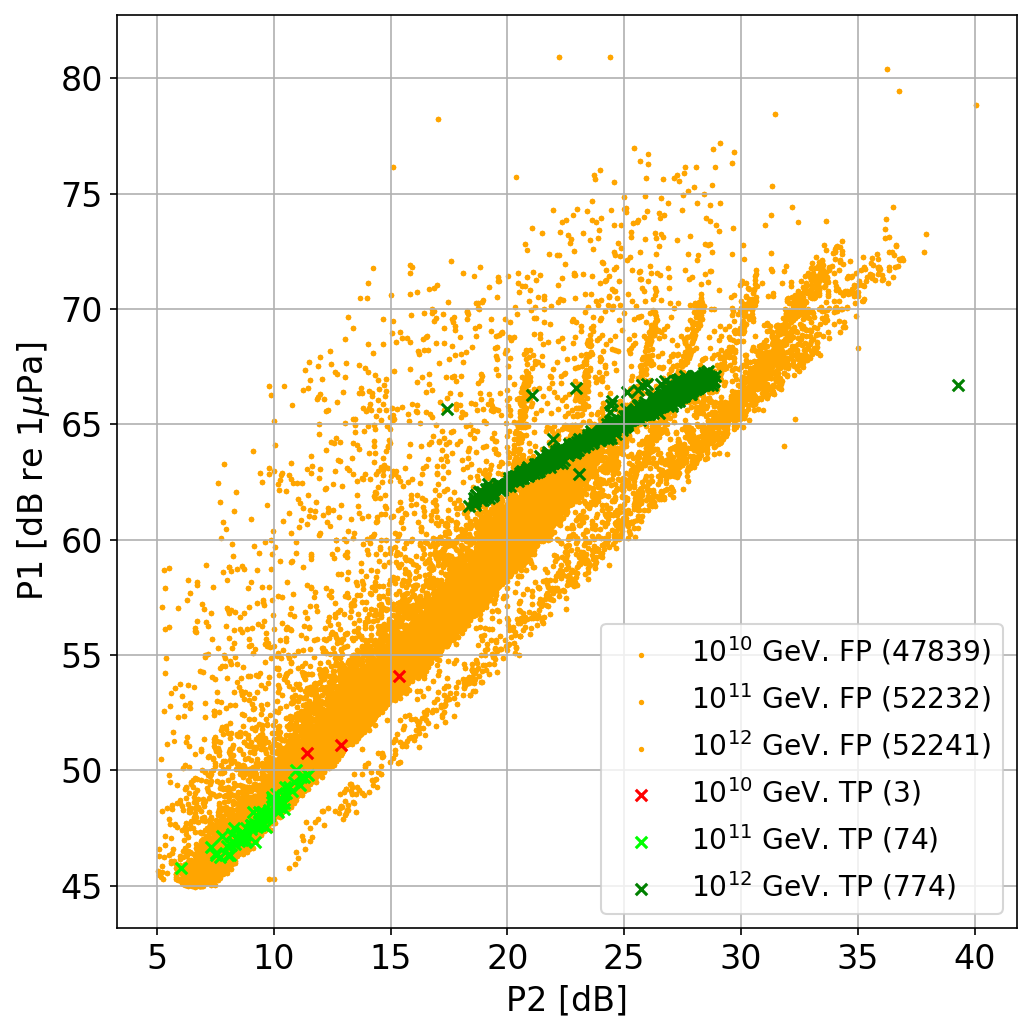}}
		\hspace{.5cm}
		\subfloat[]{\includegraphics[width=4.5cm]{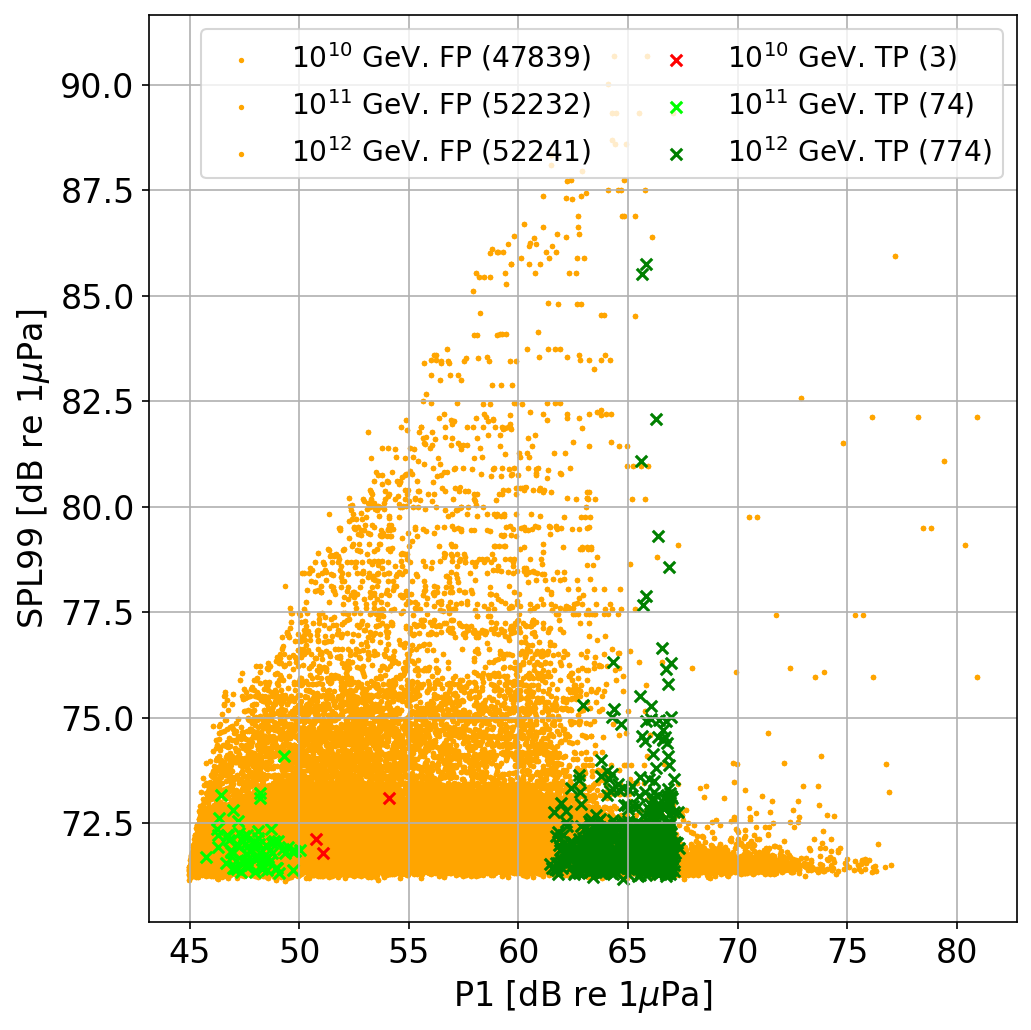}}
		\caption{TP and FP characteristics of the three different BP energies: \textbf{(a)} $P1$ vs $P1w$, \textbf{(b)} $P2$ vs $P1$, and \textbf{(c)} $P1$ vs $SPL99$.}
		\label{fig:ResGraph2}
	\end{figure}

Finally, \textit{Fig.\ref{fig:ResGraph1}} and \textit{Fig.\ref{fig:ResGraph2}} detail the characteristics of BP events detected by the algorithm. The last figure aims to show dependencies between trigger parameters. It also includes the representation of $P1$ and the 99th percentile of Sound Pressure Level (SPL) of the data portion containing the analyzed BP event, providing insight into the background noise level in impulsive noises like BPs.
    
\section{Conclusions}
This experiment acknowledges the challenge of applying the BP event trigger, which was developed and tested with hydrophones deployed at greater depths and with a sensitivity greater than 15 dB, to an O$\nu$DE-2 hydrophone. 

Based on the results obtained, it is recommended to use hydrophones with higher sensitivity and to install them in deeper water for the acoustic neutrino detection. As shown in \textit{Fig.\ref{fig:ResGraph1}}, TP events exceed the values of FP events (background) only for BPs with 10$^{12}$ GeV energy, making the detection of lower-energy events more challenging.

Although no significant change in the number of captured events is expected with the application of the second level trigger, recording the same event with four hydrophones would enable the study of the source's direction. This could lead to the development of a new criterion: if the event's direction is from top to bottom, it might be a bioacoustic event; if it is from bottom to top, it could indicate a BP produced by a neutrino interacting with the water.

Improving the temporal resolution ($T_{res.}$) of the spectrogram to achieve a bin size of 30 $\mu$s (approaching the ideal of using three samples per event) would increase the minimum detectable frequency ($F_{min.}$), potentially compromising BP detection. 

For all these reasons, acoustic neutrino detection remains one of science's greatest challenges. It is crucial to develop detectors specifically designed for this purpose rather than relying on existing technologies that do not meet the necessary requirements.

\printbibliography

@article{deFreitas2015,
  title = {Echolocation parameters of Australian humpback dolphins (Sousa sahulensis) and Indo-Pacific bottlenose dolphins (Tursiops aduncus) in the wild},
  volume = {137},
  url = {http://dx.doi.org/10.1121/1.4921277},
  DOI = {10.1121/1.4921277},
  number = {6},
  journal = {The Journal of the Acoustical Society of America},
  author = {M. de Freitas and others},
  year = {2015},
  month = jun,
  pages = {3033–3041}
}

@article{Viola2013,
  title = {NEMO-SMO acoustic array: A deep-sea test of a novel acoustic positioning system for a km3-scale underwater neutrino telescope},
  volume = {725},
  url = {http://dx.doi.org/10.1016/j.nima.2012.11.150},
  DOI = {10.1016/j.nima.2012.11.150},
  journal = {Nuclear Instruments and Methods in Physics Research Section A: Accelerators,  Spectrometers,  Detectors and Associated Equipment},
  publisher = {Elsevier BV},
  author = {S. Viola and others},
  year = {2013},
  month = oct,
  pages = {207–210}
}

@Article{Diego2022ARENA,
  author       = {D. Diego-Tortosa and others},
  date         = {2023-06},
  journaltitle = {Proceedings of Science},
  title        = {Development of a trigger for acoustic neutrino candidates in {KM3NeT}},
  doi          = {10.22323/1.424.0060},
  eprint       = {2211.10149},
  eprintclass  = {astro-ph.IM},
  url          = {https://pos.sissa.it/424/060},
  note         = {Presented in the 9th International Workshop on Acoustic and Radio EeV Neutrino Detection Activities (ARENA2022)},
  publisher    = {Sissa Medialab},
}

@PhdThesis{Diego2022Thesis,
  author      = {D. Diego-Tortosa},
  date        = {2022-09},
  institution = {Universitat Polit{\`e}cnica de Val{\`e}ncia (UPV)},
  title       = {Positioning system and acoustic studies for the {KM3NeT} deep-sea neutrino telescope},
  doi         = {10.4995/Thesis/10251/188917},
  type        = {PhD thesis},
  url         = {https://riunet.upv.es/handle/10251/188917},
  school      = {Escola Polit{\`e}cnica Superior de Gandia (EPSG)},
}

@Article{Bevan2009,
  author       = {S. Bevan and others},
  year      ={2009-05},
  journaltitle = {Nucl. Instrum. Methods Phys. Res., Sect. A},
  title        = {Study of the acoustic signature of {UHE} neutrino interactions in water and ice},
  doi          = {10.1016/j.nima.2009.05.009},
  number       = {2},
  pages        = {398-411},
  url          = {https://www.sciencedirect.com/science/article/pii/S0168900209009401},
  volume       = {607},
  journal      = {Nucl. Instrum. Methods Phys. Res., Sect. A},
  year         = {2009},
}

@article{Viola2018ARENA,
  title = {15 years of acoustic detection studies at INFN},
  volume = {216},
  url = {https://www.epj-conferences.org/articles/epjconf/abs/2019/21/epjconf_arena2018_01002/epjconf_arena2018_01002.html},
  DOI = {10.1051/epjconf/201921601002},
  note = {Presented in the 8th International Workshop on Acoustic and Radio EeV Neutrino Detection Activities (ARENA2018)},
  journal = {EPJ Web of Conferences},
  author = {S. Viola and G. Riccobene},
  year = {2019},
}

@article{Mhl2003,
  title = {The monopulsed nature of sperm whale clicks},
  volume = {114},
  ISSN = {1520-8524},
  url = {http://dx.doi.org/10.1121/1.1586258},
  DOI = {10.1121/1.1586258},
  number = {2},
  journal = {The Journal of the Acoustical Society of America},
  author = {B. Møhl and others},
  year = {2003},
  month = jul,
  pages = {1143–1154}
}

\end{document}